\documentclass[graybox]{svmult}

\usepackage{type1cm}        % activate if the above 3 fonts are
\usepackage{kotex}                        % not available on your system
\usepackage{makeidx}         % allows index generation
\usepackage{graphicx}        % standard LaTeX graphics tool
\usepackage{multicol}        % used for the two-column index
\usepackage[bottom]{footmisc}% places footnotes at page bottom
\usepackage{cite}
\usepackage{booktabs}
\usepackage{tabularx}
\usepackage{newtxtext}       % 
\usepackage[varvw]{newtxmath}       % selects Times Roman as basic font

\makeindex             % used for the subject index
\begin{document}

\title*{Optimality in group-driven social dynamics on hypergraphs}
%\author{Jihye Kim\orcidID{0009-0008-9733-2672} \\ Deok-Sun Lee\orcidID{0000-0002-5093-6582} and\\ K.-I. Goh\orcidID{0000-0003-0385-8208}}
\author{Jihye Kim \\ Deok-Sun Lee and\\ K.-I. Goh}
\institute{Jihye Kim 
\at Korea University, Seoul 02841, Korea. 
\email{arrr3755@naver.com}
\and Deok-Sun Lee \at Korea Institute for Advanced Study, Seoul 02455, Korea. \email{deoksunlee@kias.re.kr}
\and K.-I. Goh \at Korea University, Seoul 02841, Korea. \email{kgoh@korea.ac.kr}}
\maketitle

\abstract{We explore the role of intrinsic structural properties of hypergraphs in governing group-driven social dynamics with social reinforcement.  
First, we analyze simplicial contagion dynamics on random hypergraphs in which the level of hyperedge nestedness is systematically controlled. By developing the facet-based approximate master equation~(FAME) method, we demonstrate that hyperedge nestedness induces a non-monotonic change in the outbreak threshold for simplicial contagion, displaying the lowest threshold at an intermediate level of hyperedge nestedness due to competition between simple and higher-order contagion processes. 
Next, we formulate the group-driven voter model~(GVM) and investigate the consensus time for the GVM on hypergraphs with $N$ nodes. Focusing on a representative case of the GVM, we show that the consensus time scales logarithmically with the system size as $\mathcal{A}\ln N$, where the prefactor $\mathcal{A}$ displays the fastest consensus formation at an intermediate level of social reinforcement due to competition between group-constraint and nonlinearity factors. 
Taken together, our results highlight the importance of competing effects arising from higher-order interactions in shaping optimality in group-driven social dynamical processes.
}

\section{Introduction}
\label{sec:1}
Individuals in society interact within numerous social groups~(e.g. families, friend circles, clubs, and professional teams), which consist of two or more people. 
Social interactions in such groups often involve nonlinear effects~\cite{psysoc-ref4}, whereby the influence of a group on an individual is not a mere linear superposition of those of pairwise interactions. Accordingly, hypergraphs have emerged as a natural framework for modeling social systems with higher-order group interactions~\cite{phys-ref1}. Compared to graphs, hypergraphs possess two additional fundamental structural degrees of freedom that remain relatively underexplored: $(i)$ the number of nodes within a hyperedge~(called the size or cardinality of a hyperedge), and $(ii)$ the extent of overlap between distinct hyperedges. Despite a rapidly growing body of studies on dynamics on higher-order networks~\cite{phys-ref5,phys-ref6,phys-ref7,phys-ref8,phys-ref10,phys-ref11,phys-ref12,phys-my0,phys-ref14,phys-ref15,phys-ref16,phys-ref18}, there remains limited theoretical insight into how these structural features influence social dynamics~\cite{social-dynamics}. In this work, we aim to provide theoretical insight into this issue by analyzing two archetypal discrete-state dynamics:~social contagion and opinion dynamics.

\subsection{Social contagion dynamics}
Contagion is the spread of phenomena, such as diseases, behaviors, or economic crises, across biological and social systems~\cite{phys-ref54}.
The most basic models of contagion processes on graphs are the 
susceptible--infectious--susceptible~(SIS) and susceptible--infectious--recovered~(SIR) models, in which a susceptible node becomes infectious with a given transmission rate, often denoted by $\beta$, through an edge connected to an infectious neighbor. In SIS dynamics, an infectious node  returns susceptible at rate $\mu$, whereas in SIR dynamics, it becomes recovered (i.e., immune) at rate $\mu$. Because infection can occur through a single contact with an infectious neighbor, such processes are commonly referred to as {\it simple} contagion processes. 

\begin{figure}[b]
\makebox[\textwidth][c]{\includegraphics[width=0.7\textwidth]{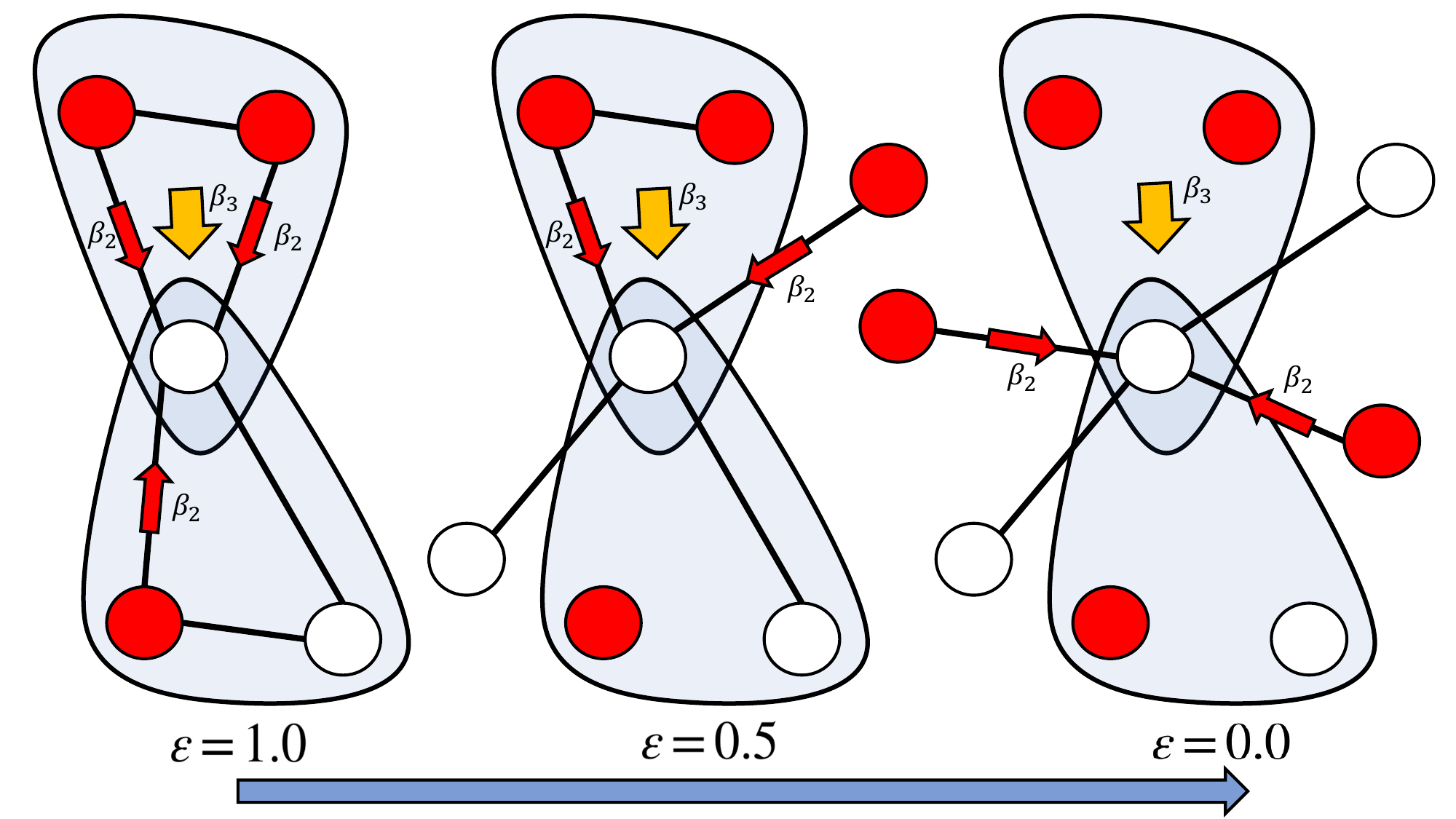}}
\caption{Schematic representation of simplicial contagion events on a specific part of an example hypergraph with hyperedge-nestedness level $\varepsilon$. The hypergraph is generated via a rewiring process from a fully nested hypergraph~(where $\varepsilon=1$). White dots correspond to susceptible nodes, whereas red dots represent infectious nodes. $\beta_2$ denotes the infection rate through a pairwise edge (solid line) and $\beta_3$ denotes that through a size-$3$ hyperedge (rounded triangle).}
\label{fig:fig0}       
\end{figure}

However, social contagion phenomena~(such as adoption of unproven new technologies) often involve social reinforcement~\cite{phys-ref42}: the cumulative social validation and perceived risk reduction arising from observing multiple neighbors who have already adopted a given idea or behavior. A common approach to model social reinforcement is to introduce a threshold mechanism, whereby contagion occurs only when the number or fraction of infectious neighbors exceeds a prescribed threshold. Motivated by both social reinforcement and the group-based nature of interactions in complex systems, Ref.~\cite{phys-ref6} has introduced the so-called {\it simplicial} contagion process~(see Fig.~\ref{fig:fig0}): a susceptible node $v$ becomes infectious through a size-$s$ hyperedge at a rate $\beta_s$ only if all the remaining $s-1$ nodes other than $v$ in that hyperedge are already infectious.
The recovery process follows that of simple contagion models defined on graphs.

In recent years, simplicial contagion models on higher-order networks have attracted considerable attention~\cite{phys-ref48, phys-ref49, phys-ref50, phys-my0, phys-ref14, contagion-dynamics-on-HONs,phys-ref52}.
The central observable is the fraction of susceptible nodes at time $t$, denoted by $S(t)$.
The stationary limit of this quantity determines the final outbreak size $R_{\infty}\equiv 1-S(t\rightarrow\infty)$. 
 Depending on the control parameters, the system ultimately settles into one of two final macroscopic states, namely, the state with outbreak ($R_{\infty}>0$) and the state without outbreak ($R_{\infty}=0$). 
The phase transition of simplicial contagion dynamics is determined not only by the degree distribution but also by the topological arrangement of hyperedges. A fundamental property of hypergraphs is {\it nestedness}: a hyperedge can be entirely contained~(that is, nested) within another larger hyperedge~\cite{phys-my0}.  

Previous studies~\cite{phys-my0, phys-ref14, phys-ref52} have reported that increasing the level of hyperedge nestedness facilitates the emergence of outbreaks for simplicial contagion dynamics on hypergraphs composed of size-$2$ and size-$3$ hyperedges.
This observation can be intuitively understood as follows:
when infection occurs through a nested pairwise edge, the condition for subsequent transmission~(i.e., infection attempt) via the associated (nesting) triangular hyperedge is concomitantly satisfied. In this sense, hyperedge nestedness promotes triangular infections, thereby making the outbreaks of simplicial contagion easier to achieve, supporting the previous reports.
Yet, this is not the end of the story: An increase in hyperedge nestedness also implies that pairwise edges become increasingly confined within triangular hyperedges, so that fewer pairwise edges remain available to serve as bridges between distinct triangular hyperedges. The overall impacts of hyperedge nestedness on final outbreak is therefore non-trivial and warrants systematic investigation. 

\begin{figure}[t]
\makebox[\textwidth][c]{\includegraphics[width=0.97\textwidth]{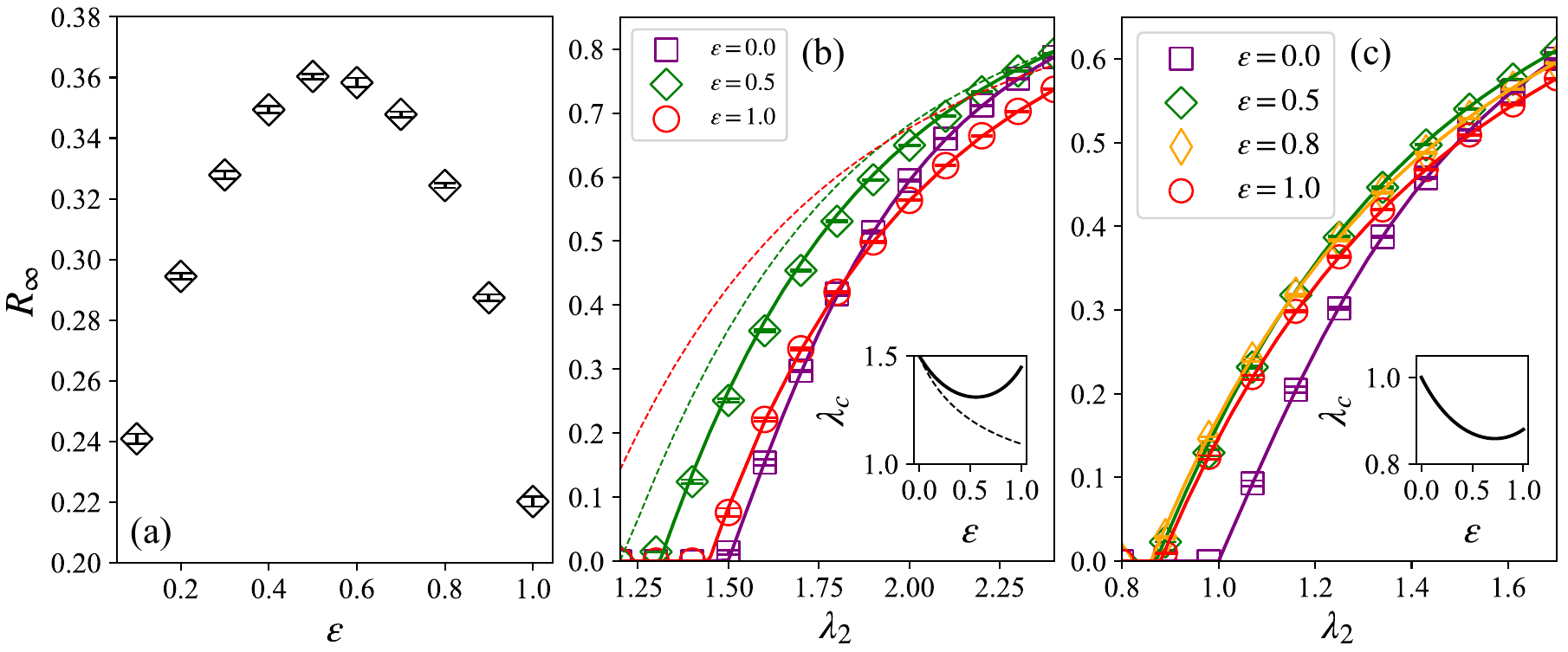}}
\caption{The outbreak size $R_{\infty}$ for simplicial SIR dynamics on random hypergraphs with joint degree distribution $P(k_2, k_3)$ and nestedness $\varepsilon$. We use the rescaled infectivity parameters $\lambda_s\equiv \langle k_s \rangle \beta_s/\mu$, where $\langle k_s \rangle = \sum_{\{k_2,k_3\}} k_s P(k_2, k_3)$. Symbols and error bars depict averages and standard errors of $R_{\infty}$ computed over $10^2$ MC simulations (using the Gillespie algorithm) on hypergraphs with $N=10^6$ nodes. We set $\lambda_3=3$. (a) $R_{\infty}$ as a function of $\varepsilon$ with $\lambda_2=1.6$ on random $(k_2=6,k_3=3)$-regular hypergraphs with $P(k_2,k_3)=\delta_{k_2,6}\delta_{k_3,3}$. Here, $\delta_{a,b}=1$ if $a=b$, and $\delta_{a,b}=0$ otherwise. (b--c) $R_{\infty}$ vs.\ $\lambda_2$ on (b) random $(k_2=6,k_3=3)$-regular hypergraphs and (c) random hypergraphs with $P(k_2,k_3)=\delta_{k_2,2k_3}\langle k_3 \rangle^{k_3} e^{-\langle k_3 \rangle}/(k_3 !)$ and $\langle k_3 \rangle =3$.
The insets show the rescaled outbreak threshold $\lambda_c$, the value of $\lambda_2$ above which the outbreak occurs, as a function of $\varepsilon$, computed analytically.
Solid lines are obtained by FAME method and dashed lines by using approximation proposed in Ref.~\cite{phys-ref52}.}
\label{fig:fig_SIR0}       
\end{figure}

Figure~\ref{fig:fig_SIR0} illustrates the results for the outbreak size $R_{\infty}$ obtained from Monte Carlo (MC) simulations of simplicial SIR dynamics using the Gillespie algorithm.
The hyperedge-nestedness parameter $\varepsilon$ denotes the average number of nested pairwise edges within a triangular hyperedges, normalized by three (Fig.~\ref{fig:fig0}).
Figs.~\ref{fig:fig_SIR0}(a--b) show the results on random $(k_2,k_3)$-regular hypergraphs, where each node belongs to $k_3$ triangular hyperedges and $k_2$ pairwise edges; here $k_3=3$ and $k_2=2k_3=6$ are used.
Fig.~\ref{fig:fig_SIR0}(a) shows that the outbreak size reaches its maximum at $\varepsilon\approx 0.5$.
Furthermore, Fig.~\ref{fig:fig_SIR0}(b) demonstrate that the outbreak threshold at which a phase transition between the two phases~($R_{\infty}=0$ and $R_{\infty}>0$) occurs, reaches its minimum at $\varepsilon\approx 0.5$. Both results indicate that the spreading becomes {\it optimal} at the intermediate level of nestedness.
Figure~\ref{fig:fig_SIR0}(c) further illustrates corresponding results on random hypergraphs with Poisson-distributed $k_3$ with $k_2=2k_3$, where the optimality remains observable in the presence of degree heterogeneity.

To analyze analytically the dynamical correlations between local dynamics through triangular hyperedges and their nested pairwise edges, we develop what we call the facet-based approximate master equation~(FAME) method (see Sec.~2 for details). 
As shown in Fig.~\ref{fig:fig_SIR0}(b--c), results of application of the FAME method (solid lines) to simplicial SIR dynamics are very accurate (main panels) and it successfully accounts for the observed optimality that the outbreak threshold becomes minimum at the intermediate nestedness level (insets). The dashed lines in Fig.~\ref{fig:fig_SIR0}(b) are obtained by using the recently proposed analytical approximation~\cite{phys-ref52}, displayed for comparison with the FAME method.
Using FAME method, we show that the optimality persists provided that the variance of the degree distribution remains below a certain threshold.

\subsection{Opinion formation dynamics}
The contagion dynamics introduced so far provide a paradigmatic framework for understanding emergent phenomena in complex systems. A closely related class of models is given by opinion formation dynamics~\cite{opinion-dynamics}.  
The standard voter model~(VM)~\cite{phys-ref23} is among the simplest graph-based models for binary opinion formation.
The binary choices are denoted by $\sigma=1$ and $\sigma=0$. At each time step, 
a uniformly random node $i$ chooses one of its pairwise edges uniformly at random, and it flips its opinion $\sigma_i$ to the opinion $\sigma_{j}$ of the neighboring node $j$ at the chosen edge if $\sigma_{j}=1-\sigma_{i}$.
The VM has been studied actively on graphs for more than two decades ~\cite{phys-ref26,phys-ref27,phys-ref28,phys-ref29,phys-ref30,phys-ref31}. Various extensions~\cite{phys-ref33} of the VM have been proposed to incorporate more realistic features of social interactions, including social reinforcement~\cite{phys-ref42,phys-ref39} and adaptive group interactions~\cite{phys-ref29,phys-ref40,phys-ref41}. However, the role of group-size constraints in shaping voter dynamics has received comparatively little attention.

\begin{figure}[b]
\sidecaption
\makebox[\textwidth][c]{\includegraphics[width=0.33\textwidth]{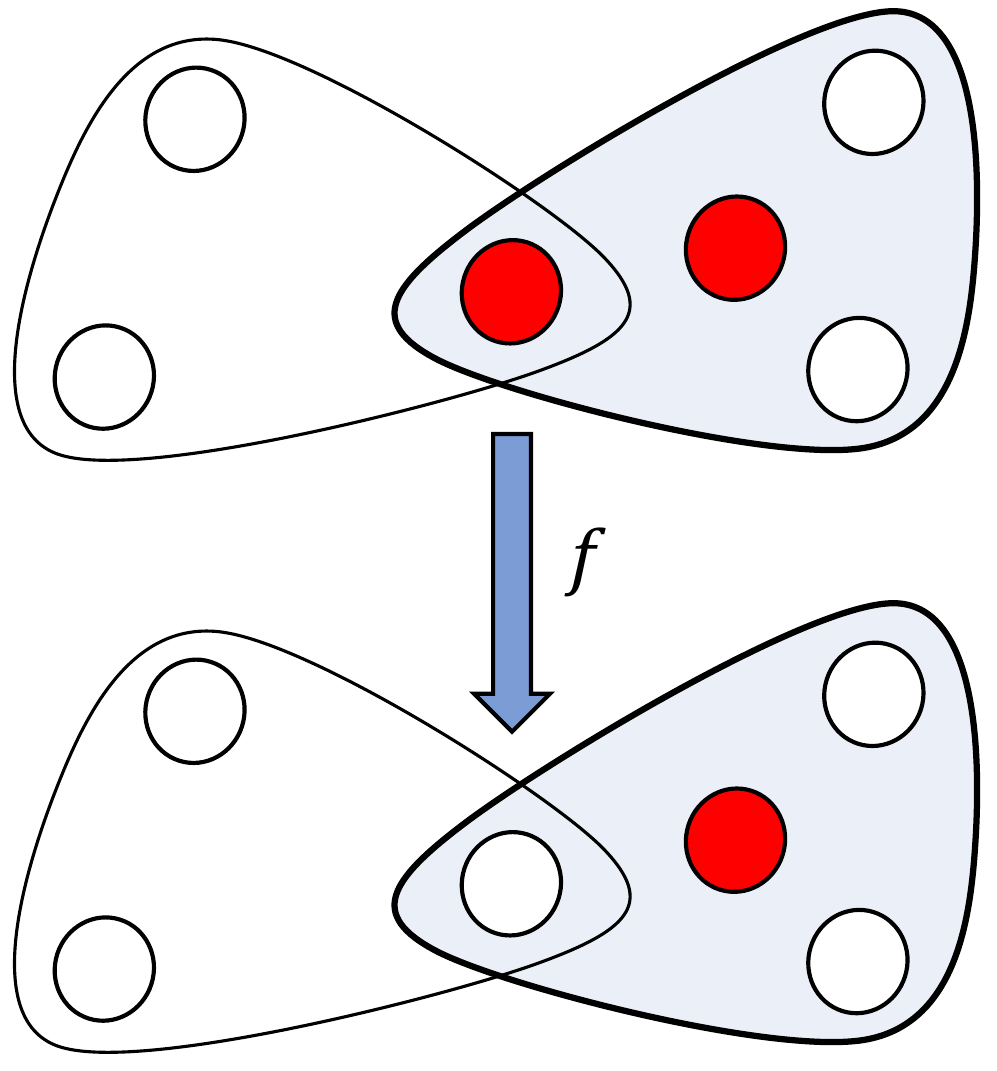}}
\caption{Schematic illustration of the group-driven voter model~(GVM) on a hypergraph. The red and white dots represent nodes holding opinion $1$ and $0$, respectively.
The central node and its right-hand-side hyperedge are selected. The central node flips its opinion via the selected hyperedge with the probability $f$ that depends on the opinion update rule. For example, under the simplicial GVM rule, the central node flips its opinion with probability $f_{\textrm{simplicial}}=0$. On the other hand, $f_q=4/9$ in the $q$-GVM with nonlinearity strength $q=2$.}
\label{fig:fig1}       
\end{figure}

To shed light on this topic, we formulate a general framework, namely, the group-driven voter model~(GVM) on hypergraphs~\cite{phys-my}. At each time, a uniformly random node $i$ chooses one of its hyperedges uniformly at random, and it flips its opinion (i.e., $\sigma_i\rightarrow 1-\sigma_i$) with probability $f$. The flip probability $f$ is determined by the number $n_i$ of neighboring nodes with opposite opinion $\sigma=1-\sigma_i$ in the selected hyperedge and by the size $s$ of that hyperedge as well as by the update rule, as illustrated in Fig.~\ref{fig:fig1}.

The GVM admits various special cases. Motivated by the simplicial contagion rule in Sec.~\ref{sec:2}, one may consider the simplicial GVM, whereby node $i$ flips its opinion if and only if all the other nodes in the selected hyperedge unanimously hold  the counter-opinion $\sigma=1-\sigma_i$. Its flip probability is given by $f_{\rm simplicial}=\delta_{n_i, s-1}$, where $\delta_{a,b}=1$ if $a=b$, and $\delta_{a,b}=0$ otherwise. In this case, the number of neighbors that node $i$ consults in the selected hyperedge is fixed to $s-1$. To relax this constraint of the simplicial GVM, we introduce another version referred to as the $q$-GVM, motivated by $q$-voter model~\cite{phys-ref39}, whereby node $i$ makes $q$ observations with replacement within the selected hyperedge and flips its opinion only when all $q$ observed opinions are equal to $1-\sigma_i$. The flip probability of $q$-GVM is given by $f_q=\left(\frac{n_i}{s-1}\right)^q$. The simplicial GVM is recovered in the limit $q\rightarrow\infty$.

\begin{figure}[t]
\sidecaption
\Description{This is Figure 2.}
\makebox[\textwidth][c]{\includegraphics[width=0.80\textwidth]{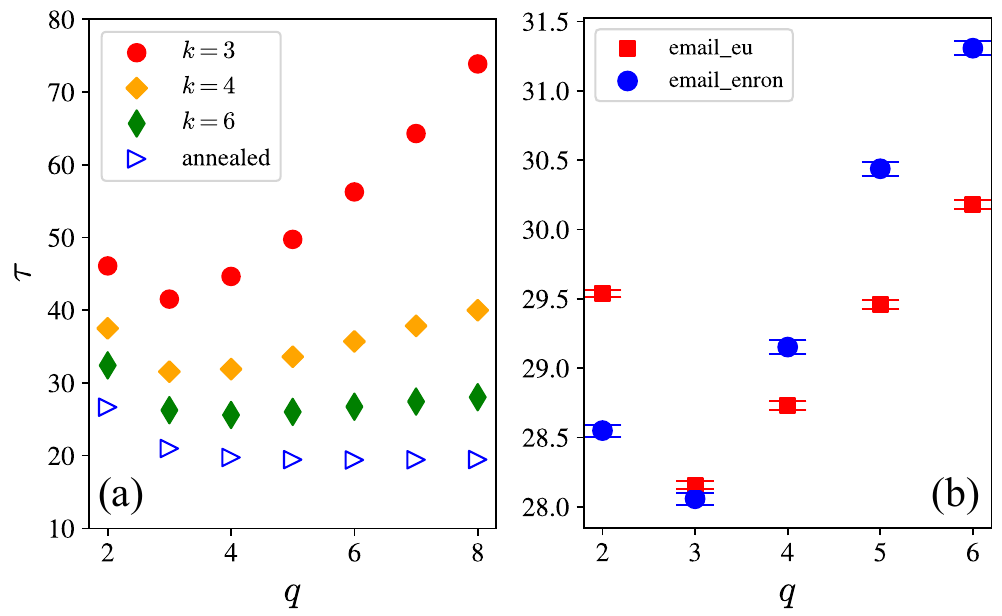}}
\caption{MC simulation results for the exit time $\tau$ of $q$-GVM on quenched hypergraphs. In (a), random $4$-uniform $k$-regular hypergraphs of $N=10^4$ nodes are used as substrates, where each hyperedge has the size $4$ and every node belongs to $k$ hyperedges. In (b), empirical hypergraphs constructed from the email communication datasets~\cite{xgi} are used. The markers (error bars) represent averages (standard errors) computed over $10^5$ independent MC simulations.}
\label{fig:fig_quenched}       
\end{figure}

In GVM, the macroscopic state of a system with $N$ nodes at time $t$ is characterized by the fraction $\rho(t)$ of nodes holding opinion $\sigma=1$. Our main quantity of interest is the consensus time, or exit time, $T(\rho_0)$, defined as the time at which $\rho(t)$ reaches $1$ or $0$ starting from an initial state with $\rho_0 \equiv \rho(0)$. In particular, we focus on the exit time $\tau\equiv T(1/2)$ from a balanced initial condition~($\rho_0=1/2$).

To initiate, let us consider the $q$-GVM on hypergraphs. Since opinion-update events would become increasingly rare as $q$ increases, it is natural to ask whether $\tau$ monotonically increases with $q$. 
Figure~\ref{fig:fig_quenched} shows the results obtained from MC simulations of the $q$-GVM on random $4$-uniform $k$-regular~[Fig.~\ref{fig:fig_quenched}(a)] and two empirical~[Fig.~\ref{fig:fig_quenched}(b)] hypergraphs. As can be seen in both panels, the consensus dynamics exhibits {\it optimality}: the exit time $\tau$ is minimized at an intermediate value of $q$. Fig.~\ref{fig:fig_quenched}(a) illustrates that this non-monotonic behavior of $\tau$ becomes less pronounced with increasing $k$. Nevertheless, this optimality persists even on annealed hypergraphs, as we shall show in Sec.~4. 

To gain analytical insight into the optimality of $\tau$, we employ mean-field theory. Within this framework, we show that $\tau$ for the $q$-GVM scales as $\mathcal{A}\ln N$, where the prefactor $\mathcal{A}$ depends both on nonlinearity $q$ and group-constraint $s$ factors. Notably, $\mathcal{A}$ depends non-monotonically on $q$ at an intermediate strength of group constraint and {\it vice versa}.
We validate the mean-field theoretical predictions for $\tau$ by means of MC simulations on annealed hypergraphs.~\footnote{The details of our MC simulations of the GVMs are provided in the Supplemental Material of Ref.~\cite{phys-my}}

\section{Simplicial contagion dynamics}\label{sec:2}
In this section, we formulate the FAME method and explore the optimality landscape for simplicial SIR and SIS dynamics on random $(k_2,k_3)$-regular hypergraphs in which every node belongs to $k_2$ pairwise hyperedges and $k_3$ triangular hyperedges with the nestedness parameter $\varepsilon$. We shall show how the optimality originates from the competition between the simple and higher-order contagion processes.

\subsection{Previous analytical methods}
In simplicial contagion dynamics, 
transmission events for a susceptible node occur via two distinct pathways: $(i)$ triangular hyperedges involving two infectious nodes, and $(ii)$ pairwise edges involving one infectious node. In the presence of nestedness, transmissions occurring through triangular hyperedges and their nested pairwise edges are not independent but dynamically correlated.   
Let us first consider the simplicial SIR dynamics on random regular hypergraphs. The fraction of nodes in state $X$~($\in\{S,I,R\}$) at time $t$ is denoted by $X(t)$.
In the homogeneous mean-field approximation~(MFA)~\cite{phys-ref6}, the time-evolution equations for $S(t)$, $I(t)$, and $R(t)$ are written as follows:
\begin{align}
\dfrac{dS}{dt} &= -(k_3 I^2 \beta_3 + k_2 I \beta_2)S, \nonumber \\
\dfrac{dI}{dt} &= -\dfrac{dS}{dt}-\mu I,
 \nonumber \\
\dfrac{dR}{dt} &= \mu I.
\label{eq:eq_mean}
\end{align}
While MFA could account for the appearance of abrupt outbreak in simplicial contagions~\cite{phys-ref6}, it by construction neglects the dynamical correlations between infection events occurring through triangular hyperedges and their nested pairwise edges. 

To capture such correlated local dynamics, 
Ref.~\cite{phys-my0} introduced the facet approximation~(FA). Therein the time-evolution equation for $S(t)$ is modified as follows:   
\begin{align}
\dfrac{dS}{dt} &= 
-\left[\Theta_{3}^{\mathrm{(free)}}(t)\beta_3+\Theta^{\mathrm{(free)}}_{2}(t)\beta_2+\Theta_2^{\mathrm{(nested)}}(t)\beta_2\right]S,
\label{eq:eq_S}
\end{align}       
in which $\Theta^{\mathrm{(free)}}_{s}(t)$ is the average number of infectious non-nested hyperedges of size $s$ per susceptible node at time $t$; $\Theta^{\mathrm{(nested)}}_{s}(t)$ is the average number of infectious nested hyperedges of size $s$ per susceptible node at time $t$. In the MFA, $\Theta_2^{\mathrm{(free)}}(t)+\Theta_2^{\mathrm{(nested)}}(t)=k_2$. The FA does not rely on the mean-field assumptions for $\Theta_3^{\mathrm{(free)}}(t)$ and $\Theta^{\mathrm{(nested)}}_{2}(t)$. Instead, FA computes them by tracking the time evolution of the probability $C^{(3)}_{n_s,n_i,n_r}(t)$ that a random triangular hyperedge contains $n_s$ susceptible nodes, $n_i$ infectious nodes, and $n_r$ recovered nodes at time $t$, in addition to the time evolution of $S(t)$ (see Table~\ref{tab:summary2}).

More recently, Ref.~\cite{phys-ref52} proposed a group-based compartmental modeling method for simplicial SIR dynamics on hypergraphs, which tracks the time evolution of the probability $\Phi^{(s)}_{m_s,m_i,m_r}(t)$ that a randomly chosen node $v$ has not yet experienced a transmission event through nodes in a randomly selected size-$s$ hyperedge which contains $m_s$ susceptible, $m_i$ infectious, and $m_r$ recovered neighboring nodes up to time $t$ ($m_s$, $m_i$, and $m_r$ satisfy the constraint $m_s+m_i+m_r=s-1$). 
This method employs the same approximation for $\Theta_2^{\mathrm{(nested)}}$ as FA and further introduces a heuristic correction term in the time-evolution equations for $\Phi^{(2)}_{m_s,m_i,m_r}(t)$. Despite the attempt to incorporate the correlated dynamics, both~\cite{phys-my0,phys-ref52} do not faithfully capture the non-monotonic effect of nestedness illustrated in Fig.~2, calling for a better analytical method.

\begin{table}[t]
\centering
\caption{A comparison of the MFA, FA, and FAME methods. Tabulated in each row are the average number $\Theta^{\mathrm{(free)}}_3(t)$ of infectious triangular hyperedges per susceptible node at time $t$, that of infectious non-nested pairwise edges~[$\Theta^{\mathrm{(free)}}_2 (t)$], and that of infectious nested pairwise edges~[$\Theta^{\mathrm{(nested)}}_2 (t)$]. 
We use the expectation value $k_2^{\mathrm{(free)}}=k_2-2k_3\varepsilon$ for the number of non-nested pairwise edges incident to a node. We also use the approximation $k_3^{(r)}\approx k_3 \binom{3}{r} \varepsilon^r (1-\varepsilon)^{3-r}$ for the number of triangular hyperedges, each containing $r$ nested pairwise edges, incident to a node which belongs to $k_3$ triangular hyperedges. Other variables are defined in the text.}
\label{tab:summary2}
\renewcommand{\arraystretch}{1.8} 
%\begin{tabularx}{\textwidth}{@{} p{3.5cm} X X X @{}}
\begin{tabularx}{\textwidth}{l XXl}
\toprule
& \textbf{MFA} & \textbf{FA} & \textbf{FAME} \\ 
\midrule
$\Theta^{\mathrm{(free)}}_3 (t)$ &
$k_3 [I(t)]^2$ &
$k_3 \dfrac{C_{1,2,0}^{(3)}(t)}{\sum\limits_{\mathbf{n'}} n'_s C^{(3)}_{\mathbf{n'}}(t) }$ &
$\sum\limits_{r}\left[ k_3^{(r)} \dfrac{\sum\limits_{\mathbf{r}} C_{1,2,0}^{(r,\mathbf{r})}(t) }{\sum\limits_{(\mathbf{n'},\mathbf{r'})} n'_s C_{\mathbf{n'}}^{(r,\mathbf{r'})}(t)} \right]$ \\
\addlinespace
$\Theta^{\mathrm{(free)}}_2 (t)$ & 
$k^{(\mathrm{free})}_2 I(t)$ & 
$k^{(\mathrm{free})}_2 I(t)$ & 
$k^{(\mathrm{free})}_2 \dfrac{C_{1,1,0}^{(2)}(t)}{\sum\limits_{\mathbf{n'}} n'_s C^{(2)}_{\mathbf{n'}}(t) }$\\ 
\addlinespace
$\Theta^{\mathrm{(nested)}}_2 (t)$~~~~~~ &
$(k_2-k^{(\mathrm{free})}_2)I(t)$ &
$k_3 \dfrac{\sum\limits_{\mathbf{n}} \varepsilon n_i n_s C_{\mathbf{n}}^{(3)}(t)}{\sum\limits_{\mathbf{n'}} n'_s C^{(3)}_{\mathbf{n'}}(t) }$ &
$\sum\limits_{r} \left[k_3^{(r)} \dfrac{\sum\limits_{\mathbf{r}} r_{si} C_{\mathbf{n}}^{(r,\mathbf{r})}(t) }{\sum\limits_{(\mathbf{n'},\mathbf{r'})} n'_s C_{\mathbf{n'}}^{(r,\mathbf{r'})}(t)} \right]$ \\
\bottomrule
\end{tabularx}
\end{table}

\subsection{Facet-based approximate master equation~(FAME) method}
To achieve a more accurate description,
we now formulate the FAME method that tracks the time evolution of the following set of probabilities:
\begin{itemize}
\item[$(i)$] ~~The probability $C^{(2)}_{\mathbf{n}}(t)$ that a random non-nested edge is in state $\mathbf{n}\equiv(n_s,n_i,n_r)$ with $n_s$ susceptible, $n_i$ infected, and $n_r$ recovered nodes, at time $t$; 
\item[$(ii)$] ~~The probability $C_{\mathbf{n}}^{(r,\mathbf{r})}(t)$ [with $\mathbf{r}\equiv (r_{si},r_{ss})$] that a random triangular hyperedge containing $r$ nested pairwise edges is in state $\mathbf{n}$ and 
includes $r_{si}$ nested S-I edges and $r_{ss}$ nested S-S edges at time $t$. 
\end{itemize}
In FAME method, $\Theta_2^{\mathrm{(free)}}(t)$ is expressed in terms of $C^{(2)}_{\mathbf{n}}(t)$;
$\Theta_3^{\mathrm{(free)}}(t)$ and $\Theta^{\mathrm{(nested)}}_{2}(t)$ are expressed in terms of $C_{\mathbf{n}}^{(r,\mathbf{r})}(t)$ (see Table~\ref{tab:summary2}).
Note that
$\mathbf{r}$ is subject to the constraints $r_{si}\leq\min\{r, n_s n_i\}$, $r_{ss}\leq\min\{r, n_s(n_s-1)/2\}$, and $r_{si}+r_{ss}\leq r$.

The time evolution of $C_{\mathbf{n}}^{(2)}(t)$ follows the master equation:
\begin{align}
\dfrac{dC^{(2)}_{\mathbf{n}}}{dt} = \mu (n_i+1)C_{n_s,n_i+1,n_r-1}^{(2)}+R^{(2)}_{n_s+1,n_i-1}(t)C^{(2)}_{n_s+1,n_i-1,n_r}
-\left[\mu n_i+R^{(2)}_{n_s,n_i}(t)\right] C^{(2)}_{\mathbf{n}}, 
\label{eq:C_2} 
\end{align}
in which $R_{n_s,n_i}^{(2)}(t)$ is the transition rate with which the state configuration of a random non-nested edge changes from $(n_s,n_i,n_r)$ to $(n_s-1,n_i+1,n_r)$ by an infection event.
It is given by $R_{n_s,n_i}^{(2)}(t)= n_s\left[\beta_2\delta_{n_s,1}\delta_{n_i,1}+\xi_2(t)\right]$, where $\xi_2(t)$ is the rate at which a random susceptible node in a random non-nested edge becomes infectious through the other hyperedges incident to the random node at time $t$. We take mean-field approximation   
for $\xi_2 (t)$ as:
\begin{align}
\xi_2 (t) = (k_2^{\mathrm{(free)}}-1)\dfrac{\Theta^{\mathrm{(free)}}_2 (t)}{k^{\mathrm{(free)}}_2}\beta_2 + \Theta_3^{\mathrm{(free)}}(t)\beta_3 + \Theta_2^{\mathrm{(nested)}}(t)\beta_2. 
\label{eq:xi2_AME}
\end{align}
We now present the time-evolution equations for $C_{\mathbf{n}}^{(r,\mathbf{r})}(t)$. For triangular hyperedges with a given $\mathbf{n}$ that contain three nested pairwise edges, we have $\mathbf{r}=(n_s n_i, \dfrac{n^2_s-n_s}{2})$. On the other hand, for the case of $r=0$, we simply have $\mathbf{r}=(0,0)$ regardless of $\mathbf{n}$.
The time-evolution equations for $C_{\mathbf{n}}^{(0,\mathbf{r})}(t)$ and $C_{\mathbf{n}}^{(3,\mathbf{r})}(t)$ are written as follows:
\begin{align}
\dfrac{dC^{(0,\mathbf{r})}_{\mathbf{n}}}{dt} &= \mu (n_i+1)C_{n_s,n_i+1,n_r-1}^{(0,\mathbf{r})}+(n_s+1)[\delta_{n_s+1,1}\delta_{n_i-1,2}\beta_3 + \xi_3^{(0)}(t)]C^{(0,\mathbf{r})}_{n_s+1,n_i-1,n_r}
\nonumber \\ &\quad-\left[\mu n_i+ n_s \delta_{n_s,1}\delta_{n_i,2}\beta_3 +n_s\xi_3^{(0)}(t)\right] C^{(0,\mathbf{r})}_{\mathbf{n}}, \\
\dfrac{dC^{(3,\mathbf{r})}_{\mathbf{n}}}{dt} &= \mu (n_i+1)C_{n_s,n_i+1,n_r-1}^{(3,\mathbf{r})}+(n_s+1)[\delta_{n_s+1,1}\delta_{n_i-1,2}\beta_3 + (n_i-1)\beta_2+\xi_3^{(3)}(t)]C^{(3,\mathbf{r})}_{n_s+1,n_i-1,n_r}
\nonumber \\ &\quad-\left[\mu n_i+ n_s \delta_{n_s,1}\delta_{n_i,2}\beta_3+n_s n_i \beta_2 +n_s\xi_3^{(3)}(t)\right] C^{(3,\mathbf{r})}_{\mathbf{n}}.
\label{eq:C_homo}
\end{align}
The master equations governing $C_{\mathbf{n}}^{(r,\mathbf{r})}(t)$ with $r=1$ and $r=2$ do not admit a simple generalized form analogous to Eqs.~(5) and~\eqref{eq:C_homo}.  
The corresponding time-evolution equations for $C_{\mathbf{n}}^{(1,\mathbf{r})}(t)$ and $C_{\mathbf{n}}^{(2,\mathbf{r})}(t)$ are given below:
\begin{align}
\dfrac{dC^{(1,(0,1))}_{3,0,0}}{dt} &= -3\xi_3^{(1)}(t) C^{(1,(0,1))}_{3,0,0}, \nonumber \\
\dfrac{dC^{(1,(0,1))}_{2,1,0}}{dt} &= \xi_3^{(1)}(t) C^{(1,(0,1))}_{3,0,0}-[\mu+2\xi_3^{(1)}(t)]C^{(1,(0,1))}_{2,1,0}, \nonumber \\
\dfrac{dC^{(1,(1,0))}_{2,1,0}}{dt} &= 2\xi_3^{(1)}(t)C^{(1,(0,1))}_{3,0,0}-[\mu+\beta_2+2\xi_3^{(1)}(t)]C^{(1,(1,0))}_{2,1,0}, \nonumber  \\ 
\dfrac{dC^{(1,(0,0))}_{1,2,0}}{dt} &= [\beta_2+\xi_3^{(1)}(t)] C^{(1,(1,0))}_{2,1,0}
-[2\mu+\beta_3+\xi_3^{(1)}(t)]C^{(1,(0,0))}_{1,2,0}, \nonumber \\ 
\dfrac{dC^{(1,(1,0))}_{1,2,0}}{dt} &= 2\xi_3^{(1)}(t) C^{(1,(0,1))}_{2,1,0}+\xi_3^{(1)}(t)C^{(1,(1,0))}_{2,1,0}-[2\mu+\beta_2+\beta_3+\xi_3^{(1)}(t)]C^{(1,(1,0))}_{1,2,0}, \nonumber \\ 
\dfrac{dC^{(1,(0,0))}_{2,0,1}}{dt} &= \mu C^{(1,(1,0))}_{2,1,0} -2\xi_3^{(1)}(t)C^{(1,(0,0))}_{2,0,1}, \nonumber \\ 
\dfrac{dC^{(1,(0,1))}_{2,0,1}}{dt} &= \mu C^{(1,(0,1))}_{2,1,0}-2\xi_3^{(1)}(t)C^{(1,(0,1))}_{2,0,1}, \nonumber \\ \displaybreak
\dfrac{dC^{(1,(0,0))}_{1,1,1}}{dt} &= 2\xi_3^{(1)}(t)C^{(1,(0,0))}_{2,0,1}+\mu C^{(1,(1,0))}_{1,2,0}+2\mu C^{(1,(0,0))}_{1,2,0}-[\mu+\xi^{(1)}_3 (t)]C^{(1,(0,0))}_{1,1,1}, \nonumber\\ 
\dfrac{dC^{(1,(1,0))}_{1,1,1}}{dt} &= 2\xi_3^{(1)}(t)C^{(1,(0,1))}_{2,0,1}+\mu C^{(1,(1,0))}_{1,2,0} -[\mu+\beta_2+\xi_3^{(1)}(t)]C^{(1,(1,0))}_{1,1,1},\nonumber \\
\dfrac{dC^{(1,(0,0))}_{1,0,2}}{dt} &= \mu(C^{(1,(1,0))}_{1,1,1}+C^{(1,(0,0))}_{1,1,1}) 
-\xi^{(1)}_3(t) C^{(1,(0,0))}_{1,0,2}; \nonumber \\
\dfrac{dC^{(2,(0,2))}_{3,0,0}}{dt} &= -3\xi_3^{(2)}(t)C^{(2,(0,2))}_{3,0,0}, \nonumber\\
\dfrac{dC^{(2,(1,1))}_{2,1,0}}{dt} &= 2\xi_3^{(2)}(t) C^{(2,(0,2))}_{3,0,0}-[\mu+\beta_2+2\xi_3^{(2)}(t)]C^{(2,(1,1))}_{2,1,0}, \nonumber \\
\dfrac{dC^{(2,(2,0))}_{2,1,0}}{dt} &= \xi_3^{(2)}(t)C^{(2,(0,2))}_{3,0,0}-[\mu+2\beta_2+2\xi_3^{(2)}(t)]C^{(2,(2,0))}_{2,1,0} \nonumber \\
\dfrac{dC^{(2,(1,0))}_{1,2,0}}{dt} &= [\beta_2+\xi_3^{(2)}(t)] C^{(2,(1,1))}_{2,1,0}+2[\beta_2+\xi_3^{(2)}(t)] C^{(2,(2,0))}_{2,1,0}-[2\mu+\beta_3+\beta_2+\xi_3^{(2)}(t)]C^{(2,(1,0))}_{1,2,0}, 
\nonumber \\
\dfrac{dC^{(2,(2,0))}_{1,2,0}}{dt} &= \xi_3^{(2)}(t) C^{(2,(1,1))}_{2,1,0}
-[2\mu+2\beta_2+\beta_3+\xi_3^{(2)}(t)]C^{(2,(2,0))}_{1,2,0}, \nonumber \\
\dfrac{dC^{(2,(0,0))}_{2,0,1}}{dt} &= \mu C^{(2,(2,0))}_{2,1,0}-2\xi_3^{(2)}(t) C^{(2,(0,0))}_{2,0,1}, \nonumber \\
\dfrac{dC^{(2,(0,1))}_{2,0,1}}{dt} &= \mu C^{(2,(1,1))}_{2,1,0}-2\xi_3^{(2)}(t) C^{(2,(0,1))}_{2,0,1},
\nonumber \\
\dfrac{dC^{(2,(1,0))}_{1,1,1}}{dt} &= 2\xi_3^{(2)}(t)C^{(2,(0,1))}_{2,0,1}+2\mu C^{(2,(2,0))}_{1,2,0}+\mu C^{(2,(1,0))}_{1,2,0}-[\mu+\beta_2+\xi^{(2)}_3(t)]C^{(2,(1,0))}_{1,1,1},
\nonumber \\
\dfrac{dC^{(2,(0,0))}_{1,1,1}}{dt} &= 2\xi_3^{(2)}(t) C^{(2,(0,0))}_{2,0,1}+\mu C^{(2,(1,0))}_{1,2,0}-[\mu+\xi_3^{(2)}(t)]C^{(2,(0,0))}_{1,1,1},
\nonumber \\ 
\dfrac{dC^{(2,(0,0))}_{1,0,2}}{dt} &= \mu(C^{(2,(0,0))}_{1,1,1}+C^{(2,(1,0))}_{1,1,1})-\xi^{(2)}_3 (t) C^{(2,(0,0))}_{1,0,2}.
\label{eq:C_inhomo} 
\end{align}
In Eqs.~(5--7), $\xi^{(r)}_3(t)$ is the rate at which a random susceptible node within a random triangular hyperedge containing $r$ nested pairwise edges becomes infectious through the other hyperedges at time $t$. As in Eq.~\eqref{eq:xi2_AME}, 
we take mean-field approximation for $\xi_3^{(r)}(t)$ as:
\begin{align}
\xi_3^{(r)} (t) = \Theta_3^{\mathrm{(free)}}(t)\beta_3 + \Theta_2^{\mathrm{(free)}}(t)\beta_2 + \Theta_2^{\mathrm{(nested)}}(t)\beta_2-\dfrac{\Theta_3^{(r,\mathrm{free})}(t)\beta_3+\Theta_2^{(r,\mathrm{nested})}(t)\beta_2}{k_3^{(r)}}, 
\label{eq:xi3_AME} 
\end{align} 
where $\Theta_3^{(r,\mathrm{free})}(t)\equiv \dfrac{k_3^{(r)}\sum\limits_{\mathbf{r}} C_{1,2,0}^{(r,\mathbf{r})}(t) }{\sum\limits_{(\mathbf{n'},\mathbf{r'})} n'_s C_{\mathbf{n'}}^{(r,\mathbf{r'})}(t)}$ and $\Theta_2^{(r,\mathrm{nested})}(t)\equiv\dfrac{k_3^{(r)}\sum\limits_{\mathbf{r}} r_{si} C_{\mathbf{n}}^{(r,\mathbf{r})}(t) }{\sum\limits_{(\mathbf{n'},\mathbf{r'})} n'_s C_{\mathbf{n'}}^{(r,\mathbf{r'})}(t)}$ are used exclusively for FAME.

In the main panels of Figs.~\ref{fig:fig_SIR0}(b--c), the solid lines for $R_{\infty}$ are obtained by
numerically integrating the FAMEs, {\it i.e.}, Eqs.~(2--8) using Euler method with $\Delta t=10^{-4}$ until $I(t)$ reaches zero.
Hereafter we apply FAME method focusing on how the rescaled outbreak threshold $\lambda_c$ is affected by the strength of hyperedge nestedness controlled by $\varepsilon$. 

\begin{figure}[b]
\makebox[\textwidth][c]{\includegraphics[width=0.97\textwidth]{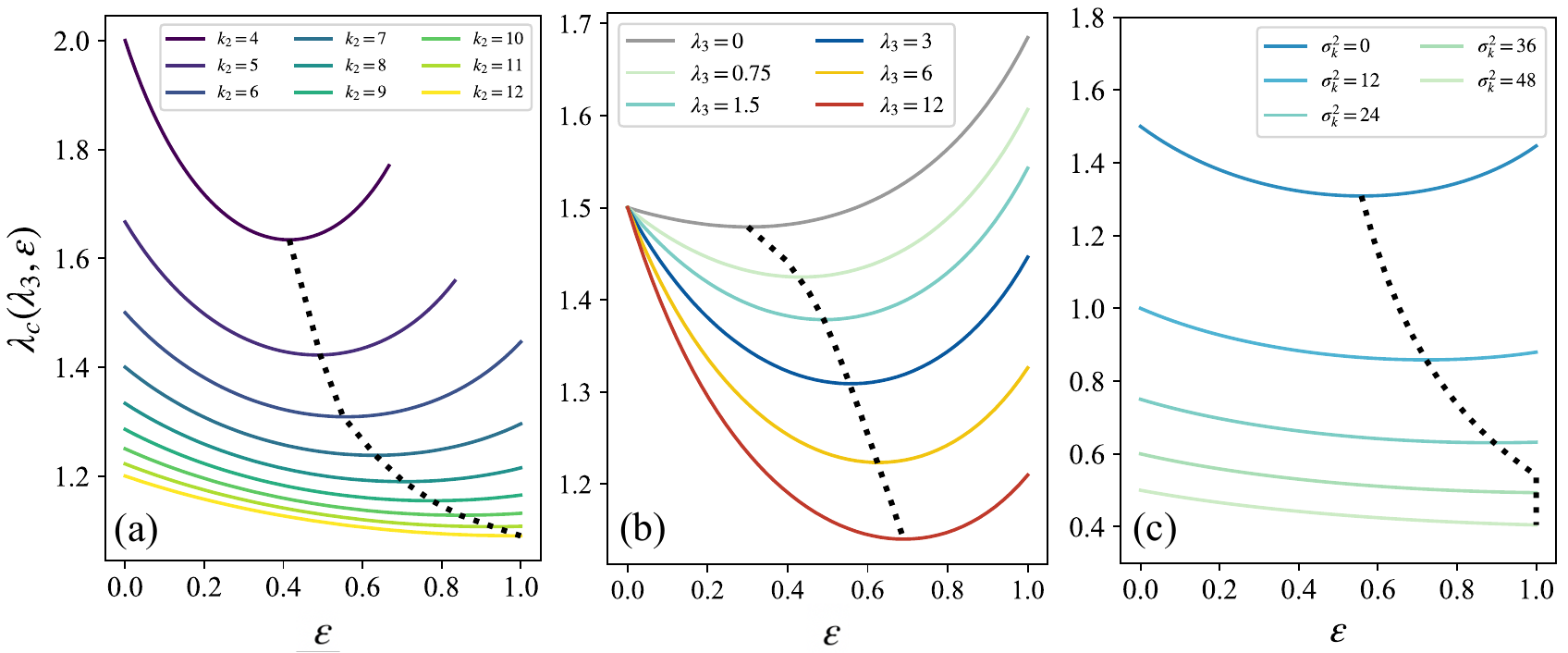}}
\caption{Rescaled outbreak threshold $\lambda_c$ for SIR dynamics on random hypergraphs with the degree distribution $P(k_2,k_3)$ and nestedness $\varepsilon$. The black dotted curves indicate the loci of $\varepsilon=\varepsilon^*$. All lines are obtained by FAME. In (a--b), the substrate hypergraphs are random $(k_2,k_3)$-regular.  
(a) $\lambda_c$  as a function of $\varepsilon$ for different values of $k_2$ when $k_3=3$ and $\lambda_3=3.0$. (b) $\lambda_c$ as a function of $\varepsilon$ for different values of $\lambda_3$ when $k_2=2k_3=6$. (c) $\lambda_c$ as a function of  $\varepsilon$ for $P(k_2,k_3)=P(k_3)\delta_{k_2,2k_3}$ with fixed mean $\langle k_2 \rangle=2\langle k_3 \rangle = 6$ and varying variance $\sigma^2_k = \langle k^2_2 \rangle -\langle k_2 \rangle^2$. It is noteworthy that $\lambda_c$ is determined by the first and second moments of $P(k_2,k_3)$, rather than by its specific form.
}
\label{fig:fig_SIR}       
\end{figure}

\subsection{Optimality of $\lambda_c$ for the simplicial contagion model}
To gain a more systematic understanding of the role of hyperedge nestedness in shaping simplicial contagion dynamics on random nested hypergraphs with the degree distribution $P(k_2,k_3)$, here we obtain the values of rescaled outbreak threshold $\lambda_c$ by evaluating the eigenvalue with the largest real part of the Jacobian matrix of Eqs.~(3,5--7) around the state $R_{\infty}=0$. For simplicial contagion dynamics on our random nested hypergraphs (in which the numbers of triangular hyperedges and pairwise edges incident to a node are correlated), $\lambda_c$ depends on $\varepsilon$, $\lambda_3$, the moments of $P(k_2,k_3)$, {\it etc}.

The insets of Figs.~\ref{fig:fig_SIR0}(b--c) have shown that $\lambda_c$ reaches its minimum at an intermediate level of hyperedge nestedness, that is, the simplicial SIR dynamics exhibits optimality. To expand on further analysis, let $\varepsilon^*$ denote the value of $\varepsilon$ that minimizes $\lambda_c$.   
Figure~\ref{fig:fig_SIR} shows the values of $\lambda_c$ and $\varepsilon^*$ obtained from the FAME method for simplicial SIR dynamics.
Figs.~\ref{fig:fig_SIR}(a--b) display $\lambda_c$ as a function of $\varepsilon$ on random $(k_2,k_3)$-regular hypergraphs. 
Fig.~\ref{fig:fig_SIR}(a) shows that for fixed $k_3=3$ with varying $k_2$, $\varepsilon^{*}$ increases with $k_2$ and eventually reaches unity at $k_{2}=k_2^*$, where $k_2^*=12$. This result suggests that the optimality against nestedness becomes effective when the propensity of size-$2$ and size-$3$ are comparable so that the spreading through the former does not overwhelm that through the latter and the two occur competitively.  In Fig.~\ref{fig:fig_SIR}(b), we find that for fixed $k_2=2k_3=6$ with varying $\lambda_3$, 
$\varepsilon^{*}$ increases with $\lambda_3$ while remaining below unity. It is noteworthy that in this case the phase transition for all $\varepsilon$ remains continuous even in the limit $\lambda_3\rightarrow\infty$.
Discontinuous transitions may emerge when, {\it e.g.}, the size of the largest facets increases or the heterogeneity in $k_3$ is sufficiently large with small $\langle k_2^2 \rangle$~\cite{phys-ref52}. 
 
The FAME framework can be extended to simplicial SIR dynamics on random non-regular hypergraphs. Unlike the regular case, nodes are classified not only by their dynamical states but also by their degree vectors $\mathbf{k}\equiv (k_2,k_3)$. For example, we can apply FAME to hypergraphs with Poisson-distributed $k_3$ and $k_2=2k_3$, the results of which was shown in Fig.~2(c). FAME works excellently when compared to the MC simulation results. In addition, Fig.~\ref{fig:fig_SIR}(c) shows how the optimality persists against the degree heterogeneity: By applying FAME for $P(k_2,k_3)=P(k_3)\delta_{k_2,2k_3}$ with fixed mean $\langle k_2\rangle=2\langle k_3\rangle=6$ but varying variance, we found that $\varepsilon^*$ increases with the variance $\sigma_k^2=\langle k_2^2\rangle-\langle k_2\rangle^2$ until it becomes unity after $\sigma_k^2\approx36$.
The results in Figs.~5(a,c) suggest that fully-nested structure can become most efficient as reported previously~\cite{phys-my0, phys-ref14, phys-ref52}, provided that the overall pairwise connectivity of hypergraphs is ensured such that the activation of higher-order contagion pathways becomes the limiting factor for global spreading.

\begin{figure}[t]
\makebox[\textwidth][c]{\includegraphics[width=0.7\textwidth]{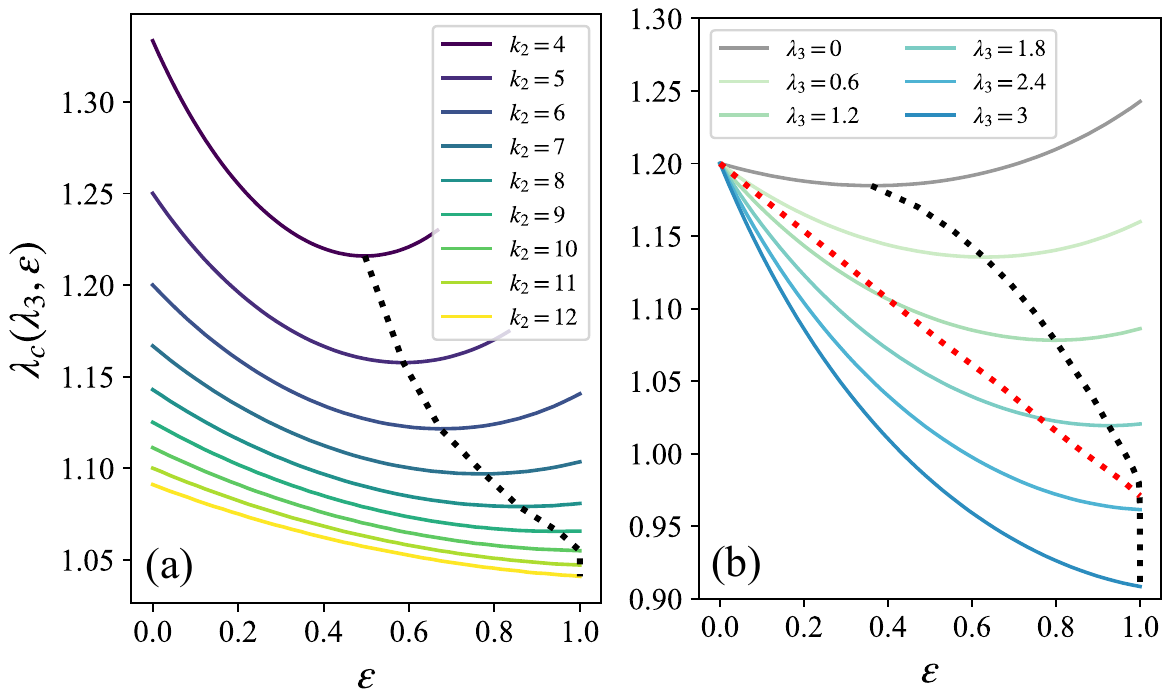}}
\caption{Rescaled outbreak threshold $\lambda_c$ for simplicial SIS dynamics on random nested $(k_2,k_3=3)$-regular hypergraphs. The black dotted curves indicate the loci of $\varepsilon=\varepsilon^*$. (a) $\lambda_c$ as function of $\varepsilon$ for different values of $k_2$ when $\lambda_3=0.75$. (b) $\lambda_c$ as a function of $\varepsilon$ for different values of $\lambda_3$ when $k_2=2k_3=6$. The red dotted curve indicates the loci of $\varepsilon=\varepsilon^{\mathrm{(tc)}}$ at which phase transitions change from discontinuous to continuous. Specifically, the phase transitions are discontinuous for $\varepsilon<\varepsilon^{\mathrm{(tc)}}$. }
\label{fig:fig_SIS}       
\end{figure}
 
 The FAME method can also be applied to SIS dynamics by replacing the terms that account for the accumulation of recovered nodes into terms that account for recovery from the infectious to the susceptible state in Eqs.~\eqref{eq:eq_S}--\eqref{eq:xi3_AME}. Simplicial SIS dynamics feature a persistent replenishment of the susceptible pool, allowing a self-sustaining reinfection cycle. As illustrated in Fig.~\ref{fig:fig_SIS}(a) for $(k_2,k_3=3)$-regular hypergraphs, the value of $\varepsilon^*$ for $\lambda_3=0.75$ remains smaller than unity for $k_2 < k^*_2=10$, mirroring the behavior of Fig.~\ref{fig:fig_SIR}(a). On the other hand, Fig.~\ref{fig:fig_SIS}(b) shows for $(6,3)$-regular hypergraphs that $\varepsilon^*$ increases with $\lambda_3$ and reaches unity at $\lambda_3\approx 2.198$, which is close to $\lambda_3\approx 2.275$ at which the phase transition becomes discontinuous even when $\varepsilon=1$.   
\footnote{Note that the prediction that optimality exists when $\lambda_3=0$ in both SIS and SIR dynamics is an artifact of the FAME method stemming from a mean-field-type approximation in 
$\Theta^{\mathrm{(nested)}}_2 (t)$. The case $\lambda_3=0$, however, can be solved without resorting to this approximation. In this case, the nested pairwise edges within triangular hyperedges for which $r<3$ can effectively be treated as non-nested pairwise edges because the only relevant correlation is among infections through triangle-forming edges. Therefore, the time-evolution equations need to be formulated only for $S(t)$, $C_{\mathbf{n}}^{(2)}(t)$ and $C_{\mathbf{n}}^{(3,\mathbf{r})}(t)$. 
In these equations, $k_2^{\mathrm{(free)}}$ is approximated by $k_2^{\mathrm{(free)}}\approx k_2-2k_3\varepsilon^3$, and we can set $k^{(r)}_3$ with $r<3$ to zero.}

The applicability of the FAME framework is constrained by a super-exponential growth in its computational complexity as the size of the largest facet increases, making a direct extension to hypergraphs with larger nested hyperedges prohibitively expensive. 
This difficulty highlights the need for reduced formulations that can nevertheless effectively capture the essential effects of hyperedge nestedness.

\subsection{Origin of optimality}
We now explain why the simplicial contagion is maximized at an intermediate value of $\varepsilon$. 
In the absence of triangular infections $(\lambda_3=0)$, an increase in $\varepsilon$ weakens the global structural connectivity of hypergraphs, thereby hampering the spread of simple contagion~(i.e., $\varepsilon^*=0.0$). 
\begin{figure}[t]
\sidecaption
\Description{This is Figure 1.}
\makebox[\textwidth][c]{\includegraphics[width=0.5\textwidth]{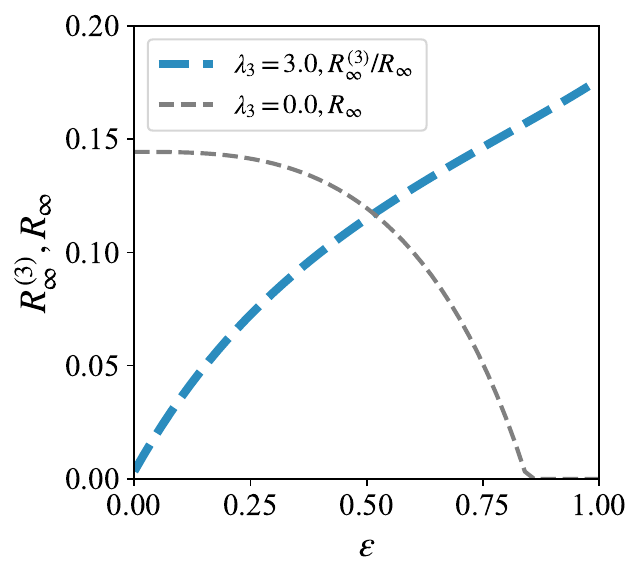}}
\caption{The ratio $R_{\infty}^{(3)}/R_{\infty}$ of outbreaks occurring through higher-order contagion events, and the total outbreak size $R_{\infty}$ for simplicial SIR with $\lambda_2=1.6$. We consider random regular hypergraphs with $k_3=3$ and $k_2=6$. Both curves are obtained from the FAMEs. 
The gray dashed line represents $R_{\infty}$ for $\lambda_3=0.0$, and the blue solid line represents $R_{\infty}^{(3)}/R_{\infty}$ when $\lambda_3=3.0$.}
\label{fig:R_inf}       
\end{figure}
This is confirmed by the gray dashed line in Fig.~\ref{fig:R_inf} for the total outbreak size $R_{\infty}$ as a function of $\varepsilon$ when $\lambda_3=0$, showing that simple contagion spreads less with increasing $\varepsilon$~(i.e., $\varepsilon^*=0$). For simple contagion dynamics driven solely by pairwise interactions, the spreading is hindered as the number of edges forming triangles increases, i.e., as the clustering coefficient increases~\cite{phys-ref55,phys-ref56}. Physical insight behind the observation $\varepsilon^*=0$ when $\lambda_3=0$, might be that an increase in $\varepsilon$ leads to a decrease in the number of paths between triangular groups, limiting the global spread of simple contagion. 

On the other hand, when $\lambda_3>0$, simple contagion through nested pairwise edges can immediately trigger higher-order contagion. To quantify this effect, we 
use FAME method to compute the ``triangle-driven oubreak size'' $R^{(3)}_{\infty}$, given by the sum of $\Theta^{\mathrm{(free)}}_{3}(t)\beta_3 S(t)$ over time. 
The blue solid line in Fig.~7 shows the ratio $R_{\infty}^{(3)}/R_{\infty}$, indicating that indeed higher-order contagion gets more activated with increasing $\varepsilon$. 
The two effects compete and there is a sweet spot $\varepsilon$ that maximizes the spread of contagion, i.e., minimizes $\lambda_c$ between the two extremes $\varepsilon=0$ and $\varepsilon=1$.

We now discuss why $\varepsilon^*$ increases with $\langle k_2 \rangle$, $\langle k_2^2 \rangle$, and $\lambda_3$. As $\langle k_2 \rangle$ and $\langle k^2_2 \rangle$ increases, the benefit of small $\varepsilon$ diminishes, given that hypergraphs become well-connected even for large $\varepsilon$. The disadvantage of small $\varepsilon$, i.e., the lack of higher-order contagion events, therefore increasingly dominates. As a result, $\varepsilon^*$ shifts to a higher value with increasing $\langle k_2 \rangle$ and $\langle k^2_2 \rangle$. 
Turning to the impact of $\lambda_3$ on $\varepsilon^*$, as $\lambda_3$ increases, the benefit of increasing $\varepsilon$, i.e., the easier satisfaction of the 
condition for higher-order contagion, gradually outweighs the benefit of decreasing $\varepsilon$, i.e., improved hypergraph connectivity.

\section{Group-driven voter model (GVM)}\label{sec:3}
In this section, we present theoretical analysis attempting to examine how the optimality in GVMs is shaped by group effects. Here, optimality refers to the existence of an intermediate nonlinearity strength that minimizes the exit time (see Fig.~\ref{fig:fig_quenched}). We summarize the analysis presented in Ref.~\cite{phys-my}, to which interested readers are referred for details.

\subsection{Mean-field theory}
The time evolution of the fraction $\rho(t)$ of nodes holding opinion $1$ follows the rate equation $\frac{d\rho}{dt}=R(\rho)-L(\rho)\equiv v(\rho)~$, where $R(\rho)$ [$L(\rho)$] denotes the probability that a uniformly random node $i$ with $\sigma_i=0$ ($\sigma_i=1$) flips its opinion.
The recursion equation for the exit time $T(\rho)$ is given by:
\begin{align}
T(\rho) = R(\rho)\,T(\rho+\delta\rho) + L(\rho)\,T(\rho-\delta\rho) + [1-R(\rho)-L(\rho)]\,T(\rho)+\delta t~,
\label{eq:T}
\end{align}
where $\delta\rho=\delta t=1/N$. 
By Taylor-expanding $T(\rho\pm \delta\rho)$ up to second order in $\delta\rho$, we obtain the backward Kolmogorov equation  
\begin{align}	
v(\rho)\dfrac{\partial{T}}{\partial{\rho}}+D(\rho)\dfrac{\partial^2{T}}{\partial{\rho^2}}&=-1 \, , \label{eq:kolmos} 
\end{align}
with $v(\rho)\equiv R(\rho)-L(\rho)$ and $D(\rho)\equiv[R(\rho)+L(\rho)]/(2N)$.  
Under the mean-field approximation,  $R(\rho)$ can be written as follows~\cite{phys-ref30,phys-my}:
\begin{align}
	R(\rho) &= \dfrac{(1-\rho)}{\sum_{s=2} sP(s)}\sum_{s=2} sP(s)\sum_{n=1}^{s - 1} \mathcal{B}(s-1,\rho,n) f(s,n) \, .  \label{eq:R}
\end{align}
Here $P(s)$ denotes the probability that a uniformly random hyperedge has size $s$; $\mathcal{B}(s-1,\rho,n)\equiv \frac{(s-1)!}{n!(s-1-n)!}\rho^{n} (1-\rho)^{s-1-n}$ is the binomial factor; 
$f(s,n)$ is the probability that a random node flips by referring to a hyperedge accommodating $s-1$ neighboring nodes among which $n$ nodes hold opinion $1$. By symmetry, $L(\rho)=R(1-\rho)$.

To obtain the exit time, one can numerically solve the recursion equation \eqref{eq:T}.\footnote{See the Supplemental Material of Ref.~\cite{phys-my} for details.} One can also obtain the approximate exit time, correct up to leading-order scaling, by neglecting the second~({\it i.e.}, diffusion) term in Eq.~\eqref{eq:kolmos}~\cite{phys-ref44,phys-my}. Specifically, we have 
\begin{align}
\tau\approx \int_{\frac{1}{2}-\frac{1}{\sqrt{N}}}^{\frac{1}{N}} \dfrac{1}{v(\rho')} d\rho'   \sim \left(-C_1 + \dfrac{C_2}{4}\right)\ln N \, ,
\label{eq:generalized_log}
\end{align}
with $C_1 =\left.\frac{\rho}{v(\rho)}\right|_{\rho=0}$ and 
$C_2 =\left.\frac{2\rho-1}{v(\rho)}\right|_{\rho=\frac{1}{2}}$.
In the next subsections, we apply Eq.~\eqref{eq:generalized_log} to the simplicial GVM and the $q$-GVM, respectively, on annealed hypergraphs.

\begin{figure}[t]
\sidecaption
\Description{This is Figure 2.}
\makebox[\textwidth][c]{\includegraphics[width=0.7\textwidth]{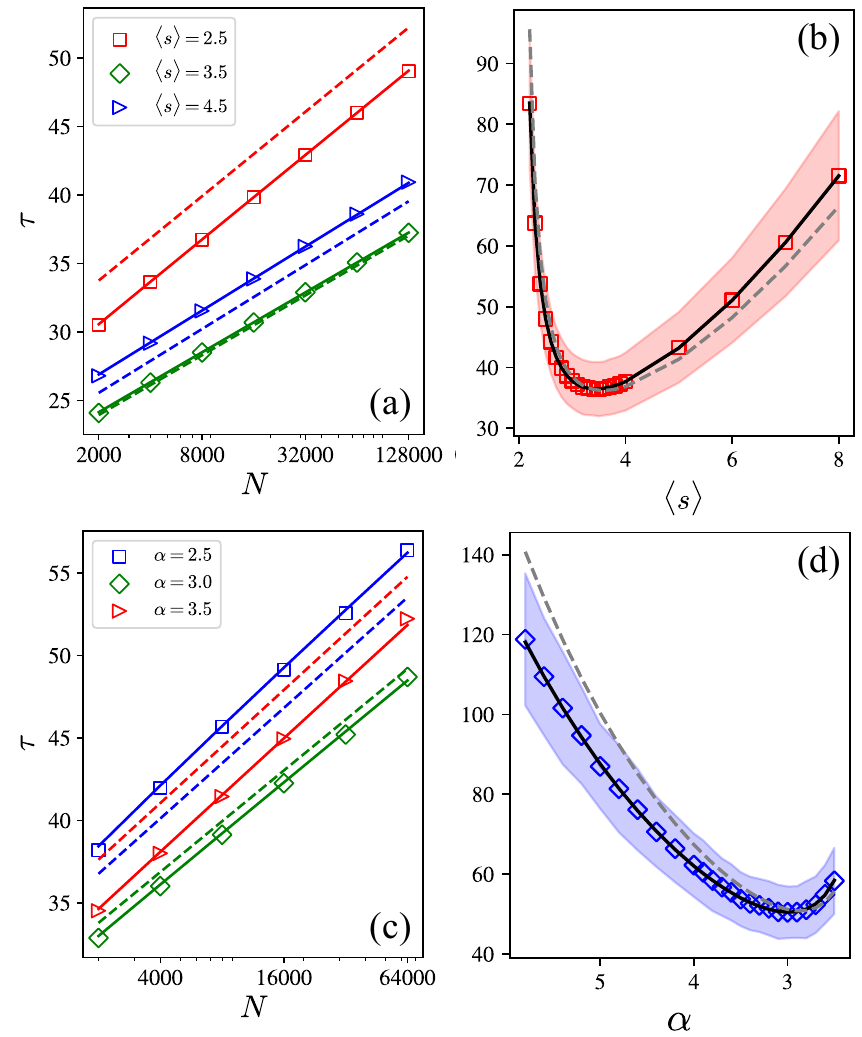}}
\caption{Exit time $\tau$ for the simplicial GVM on annealed hypergraphs with (a--b)~a geometric $P(s)$ with different mean hyperedge size $\langle s \rangle$  and (c--d)~a power law $P(s)$ with different power law exponents $\alpha$. The symbols and shaded areas indicate the averages and standard deviations, respectively, of MC simulations. The solid and dashed curves correspond to numerical solution of Eq.~\eqref{eq:T} and leading-order approximate solution of Eq.~\eqref{eq:generalized_log}, respectively. The panels (b) and (d) are reproduced from Ref.~\cite{phys-my}.}
\label{fig:fig2}      
\end{figure}

\subsection{Logarithmic scaling and optimality of $\tau$ for the simplicial GVM}
Let us first consider the simplicial GVM for which node $v$ flips its opinion $\sigma_v$ only when all $s-1$ other nodes in the randomly selected hyperedge (of size $s$) unanimously hold the opinion $\sigma=1-\sigma_v$ and thus $f(s,n)=\delta_{n,s-1}$. In this case, the logarithmic scaling of $\tau$ is given by: 
\begin{align}
\tau\sim \left[\frac{\langle s \rangle}{\langle s \rangle-2P(s)}+\dfrac{\langle s \rangle}{\sum\limits_{s=3} s(s-2)\left(\frac{1}{2}\right)^{s-3} P(s)}\right]\ln N.
\label{eq:simplicialGVM}
\end{align}
In Fig.~\ref{fig:fig2}, we present the consensus time $\tau$ of the simplicial GVM on annealed hypergraphs for two choices of hyperedge-size distribution
$P(s)$, both motivated by empirical data~\cite{phys-ref45,phys-ref46,phys-ref47}.  
Fig.~\ref{fig:fig2}(a) shows the logarithmic scaling of $\tau$ with respect to system size~$N$ for geometric hyperedge-size distribution $P(s)=\frac{1}{\langle s\rangle-1}\left(\frac{\langle s\rangle-2}{\langle s\rangle-1}\right)^{s-2}$. The theoretical predictions are consistent with the MC simulations. Fig.~\ref{fig:fig2}(b) shows the optimality of $\tau$:  
Increasing the mean hyperedge size $\langle s \rangle$ accelerates consensus formation only up to a point~($\langle s\rangle^*\approx 3.58$), beyond which it slows down. 
Figs.~\ref{fig:fig2}(c--d) show the corresponding results for power-law hyperedge-size distribution 
$P(s) = \frac{s^{-\alpha}}{\zeta(\alpha)-1}$, where $\zeta(\alpha)$ is Riemann zeta function. For this choice of $P(s)$, $\tau$ exhibits a minimum at $\alpha^* \approx 2.87$.

\subsection{Logarithmic scaling and optimality of $\tau$ for the $q$-GVM}
We now analyze the $q$-GVM, for which node $v$ flips its opinion $\sigma_v$ only when $q$ observations (with replacement) within the selected hyperedge (of size $s$) indicate unanimously the opposite opinion $\sigma=1-\sigma_v$ and thus
$f(s,n)=\left(\frac{n}{s-1}\right)^q$.
For simplicity, we focus on the $q$-GVM on annealed $s$-uniform hypergraphs with the hyperedge-size distribution $P(s')=\delta_{s',s}$. 
The logarithmic scaling of $\tau(s,q,N)$ in 
Eq.~\eqref{eq:generalized_log} is given by\footnote{The detailed derivation of  Eqs.~\eqref{eq:simplicialGVM}--\eqref{eq:sN} can be found in the Supplemental Material of Ref.~\cite{phys-my}.}
\begin{align}
\tau(s,q,N)\sim \left[\frac{(s-1)^{q-1}}{(s-1)^{q-1}-1}+ \frac{(s-1)^{q-1}}{\sum\limits_{r=1}^{q-1} rA_{r,q}\left(\frac{1}{2}\right)^{r-1} \prod_{l=1}^{r} (s-1-l)}\right]\ln N \ \, .
\label{eq:eq_qGVM}
\end{align}
with positive integers $A_{r,q} \equiv \frac{(r+1)^{q-1}-r^{q}+r-1}{r!}+\mathbf{1}_{r\geq 3}\sum\limits_{l=2}^{r-1}\frac{(r-l+1)^{q-1}-1}{l!(r-l)!}(-1)^l$, where $\mathbf{1}_{r\geq 3}$ is the indicator symbol. 

\begin{figure}[t]
\sidecaption
\makebox[\textwidth][c]{\includegraphics[width=0.7\textwidth]{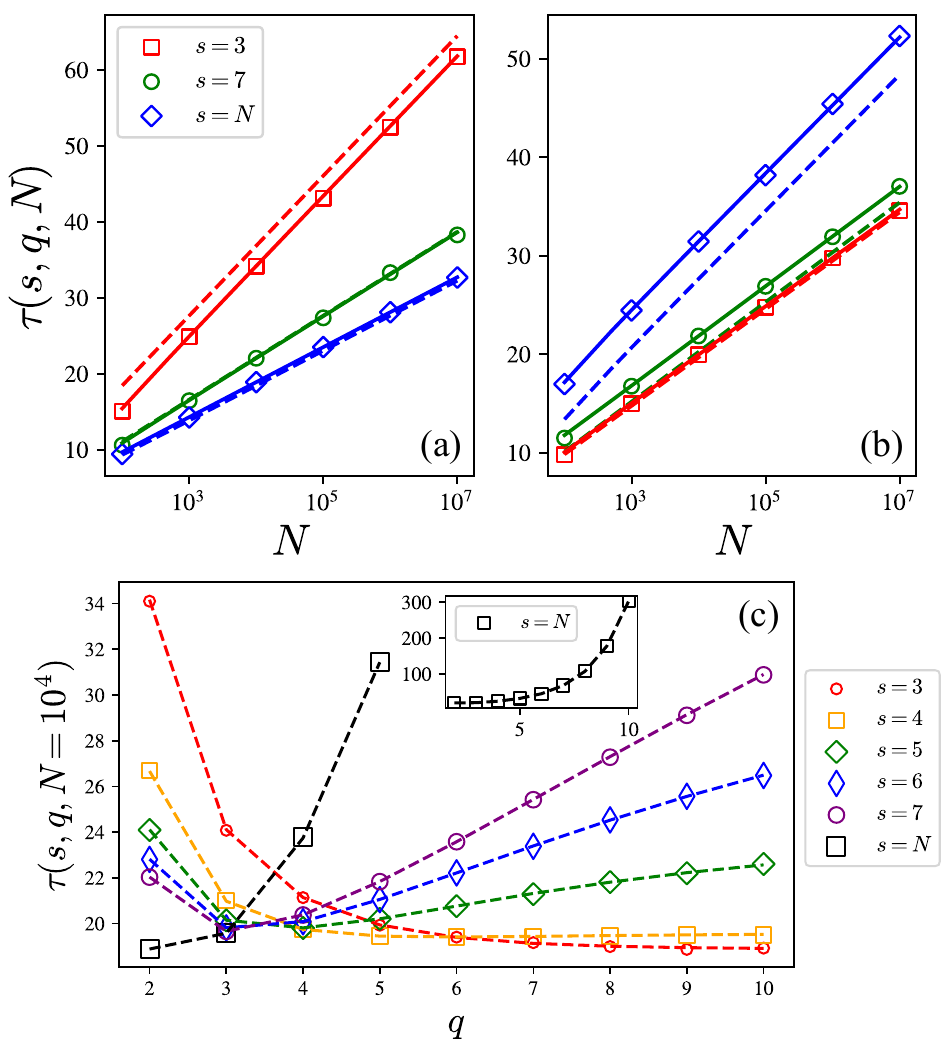}}
\caption{Exit time $\tau(s,q,N)$ for the $q$-GVM on annealed $s$-uniform hypergraphs of $N$ nodes. (a--b) Logarithmic scaling of $\tau$ against $N$ when (a) $q = 2$ and (b) $q = 5$. The markers are from MC simulations. 
Solid and dashed curves obtained from numerical solution of Eq.~\eqref{eq:T} and the leading-order solution of Eq.~(12) or (14), respectively.
(c) The dependence of $\tau$ on $q$ for different values of $s$ with $N = 10^4$. Dashed curves depict the numerical solution of the recursion relation, Eq.~\eqref{eq:T}.
In the inset, we show the curve for $s = N$ in an extended $q$ range. The panels (a--c) are reproduced from Ref.~\cite{phys-my}.}
\label{fig:fig4}       
\end{figure}

Figures~\ref{fig:fig4}(a--b) show that the analytically obtained logarithmic scaling of $\tau$ agrees well with the MC simulation results, and indicate that the effects of group sizes vary with the nonlinearity strength $q$. 
Specifically, in Fig.~\ref{fig:fig4}(a) where $q=2$ (observing two opinions), the consensus time $\tau$ decreases monotonically with $s$ and thus attains its maximum under the strongest group constraint, i.e., $s=3$. On the other hand, for $q=5$ shown in Fig.~\ref{fig:fig4}(b), $\tau$ is maximized in the opposite limit of $s=N$, where group constraint is absent. This suggests that, for some intermediate values of $q$~(e.g., $q=4$ or $q=5$), the shortest consensus time is achieved at an intermediate group size $s$.

To substantiate the existence of optimality and further examine how the nonlinearity and higher-order interactions
jointly shape the consensus formation in $q$-GVM dynamics, we investigate $\tau$ as a function of $q$ for specified values of $s$.
Two limits of group constraints in Eq.~\eqref{eq:eq_qGVM} admit simplification: When $s=N$ (group constraint is absent) and $s=3$ (strongest group constraint), we have respectively 
\begin{align}
	\tau(s=N,q,N)\sim \left(1+\dfrac{2^{q-2}}{q-1}\right)\ln N \quad\textrm{;}\quad 	\tau(s=3,q,N)&\sim\left(\dfrac{2^{q}}{2^{q-1}-1}\right)\ln N~.
\label{eq:sN}
\end{align}
As a function of $q$, the former increases, while the latter decreases, monotonically. For intermediate group sizes, neither too large nor too small $s$, optimality emerges:
Fig.~\ref{fig:fig4}(c) shows that for $s>3$ there exists optimal nonlinearity strength $2<q^*<\infty$ at which $\tau$ reaches its minimum. Conversely, for a fixed $q$, $\tau$ is minimized at an intermediate group size $s^*$, indicating the existence of an optimal group constraint that is neither too weak nor too strong~\cite{phys-my}.

\subsection{Origin of optimality}
We now provide an explanation of the observed optimality of $\tau$.
The dynamics of consensus formation becomes diffusive, yielding $\tau\sim O(N)$, in two limiting regimes. Firstly, when $s=2$ or $q=1$, for which a node consults only one neighbor, the $q$-GVM becomes linear. Secondly, if the group constraint is weak~(i.e., the value of $s$ is greater enough), increasing $q$ rapidly reduces the probability that all consulted neighbors share the same opinion. For instance, when $s=N$, if $q$ increases to logarithmic order in $N$, $q\sim O(\ln N)$, the logarithmic scaling of $\tau$ eventually crosses over to a linear scaling, $\tau\sim O(N)$, comparable to the diffusive behavior observed for the linear GVM dynamics. Between these two diffusive, slow regimes, the drift remains appreciable, driving to faster consensus formation, i.e., logarithmic scaling of $\tau$. Consequently, $\tau$ exhibits a minimum at intermediate values of $s$ and $q$ between the two limits, as shown in Fig.~\ref{fig:fig4}(c). The case of $s=3$ is an exception: The number of distinct neighbors that can be consulted is either one or two, and the probability of the latter increases with $q$, leading to $\tau$ decreasing monotonically with increasing $q$~(i.e. $q^*=\infty$). 

The intuition based on the competition between group constraint and nonlinearity explains the emergence of optimality of $\tau$ for the simplicial GVM shown in Fig.~\ref{fig:fig2}. 
In this case, the diffusive dynamics driven by pairwise hyperedges~(of size $2$) dominates as $\langle s \rangle \rightarrow 2$ and $\alpha\rightarrow\infty$. The other diffusive dynamics arising from the small probability of unanimity in large hyperedges dominates as $\langle s \rangle \rightarrow \infty$ and $\alpha\rightarrow 2$. As the system approaches either limit, $\tau$ increases, resulting again in an optimal value at intermediate parameters.

\section{Discussion}
In this work, we have examined how group-level nonlinear interactions lead to non-monotonic behavior, which we refer to as optimality, by studying two classes of group-driven social dynamics defined on hypergraphs.

We first studied on simplicial contagion dynamics, which captures the coexistence of pairwise simple contagion events and higher-order complex contagion events, on nested hypergraphs with hyperedges of size up to three. We found that simplicial contagion spreads most efficiently at an intermediate level $\varepsilon=\varepsilon^*$ of hyperedge nestedness.
To analyze how $\varepsilon^*$ depends on the hypergraph connectivity and the triangular infection rate, we introduced a novel mathematical framework, the facet-based approximate master equation~(FAME) method, which improves over existing analytical methods~\cite{phys-ref6,phys-my0,phys-ref52}. Using FAME, we
uncovered a non-monotonic dependence of rescaled outbreak threshold $\lambda_c$ on $\varepsilon$, revealing that 
the optimal nestedness $\varepsilon^*$ increases with $\langle k_2 \rangle$, $\langle k_2^2 \rangle$, and $\lambda_3$. 
The optimality of $\lambda_c$ can be attributed to a competition between two opposing effects: increasing $\varepsilon$ weakens the overall pairwise connectivity of the hypergraph, while concomitantly enhancing the activation of higher-order contagion pathways.

We then considered a minimal yet informative generalization of the voter model on hypergraphs incorporating within-group nonlinear interactions, termed the group-driven voter model~(GVM). 
For a representative case of $q$-GVM, MC simulations on annealed hypergraphs and mean-field analysis for the exit time $\tau$ revealed that consensus formation is jointly shaped by the strength of nonlinearity and group constraint. Specifically, $\tau$ scales logarithmically with $N$, i.e., $\tau\sim \mathcal{A}\ln N$ and its prefactor $\mathcal{A}$ depends on both the nonlinearity strength $q$ and the group-constraint factor $s$. 
Notably, the consensus state is achieved most rapidly 
at an intermediate group effect. The optimality of $\tau$ can be attributed to a competition between two opposing effects: increasing $q$ and $s$ suppress diffusive fluctuations, thereby promoting consensus formation, while at the same time delays drift-induced ordering, thereby hindering consensus formation. 

Overall, we have demonstrated that, within the scope of the present models, the interplay between in-group nonlinear interactions and their structural characteristics can lead to unexpected optimality in social dynamics on hypergraphs. 
We hope that our results presented in this paper could contribute to a deeper understanding of collective dynamics in complex social systems and stimulate several promising directions for future research.

\begin{acknowledgement}
We thank B.~Min, M.~A.~Porter, and M.~San Miguel for valuable collaboration on GVM. This work was supported in part by National Research Foundation of Korea~(NRF) grant funded by the Korea government~(MSIT)~(Grant No.~RS-2025-00558837)~(K.-I.G.) and by a KIAS individual Grant No.~CG079902 at Korea Institute for Advanced Study~(D.-S.L.).               
\end{acknowledgement}

\ethics{Competing Interests}{ 
The authors have no conflicts of interest to declare that are relevant to the content of this paper.}

%\eject

\ethics{Ethics Approval}{
Ethical approval was not required for this study as it did not involve human or animal subjects.}

\end{document}